%% file: paper.tex
\documentclass[11pt,a4paper]{article}
\usepackage{ifthen}
\usepackage{booktabs}
\usepackage{overpic}
\usepackage{relsize}
\usepackage{rotating}
\newboolean{pdflatex}
\setboolean{pdflatex}{true}

\newboolean{articletitles}
\setboolean{articletitles}{true}

\newboolean{inbibliography}
\setboolean{inbibliography}{false}

\usepackage{microtype}
\usepackage{lineno}
\usepackage{xspace}
\usepackage{caption}
\renewcommand{\captionfont}{\small}

\usepackage{graphicx}
\usepackage{color}
\usepackage{colortbl}

\usepackage{amsmath}
\usepackage{amssymb}
\usepackage{amsfonts}
\usepackage{upgreek}

\newcommand*\patchAmsMathEnvironmentForLineno[1]{%
\expandafter\let\csname old#1\expandafter\endcsname\csname #1\endcsname
\expandafter\let\csname oldend#1\expandafter\endcsname\csname
end#1\endcsname
 \renewenvironment{#1}%
   {\linenomath\csname old#1\endcsname}%
   {\csname oldend#1\endcsname\endlinenomath}%
}
\newcommand*\patchBothAmsMathEnvironmentsForLineno[1]{%
  \patchAmsMathEnvironmentForLineno{#1}%
  \patchAmsMathEnvironmentForLineno{#1*}%
}
\AtBeginDocument{%
\patchBothAmsMathEnvironmentsForLineno{equation}%
\patchBothAmsMathEnvironmentsForLineno{align}%
\patchBothAmsMathEnvironmentsForLineno{flalign}%
\patchBothAmsMathEnvironmentsForLineno{alignat}%
\patchBothAmsMathEnvironmentsForLineno{gather}%
\patchBothAmsMathEnvironmentsForLineno{multline}%
\patchBothAmsMathEnvironmentsForLineno{eqnarray}%
}

\usepackage{hyperref}
\usepackage[all]{hypcap}

 \mathchardef\PLambda="7103

\usepackage{cite}
\usepackage{mciteplus}

\usepackage{longtable}
\usepackage{chngpage}

\def \smdecay {\ensuremath{B^0 \!\to K^{*0}\mu^+\mu^-}\xspace}
\def \sigdecay {\ensuremath{B^0 \!\to K^{*0}\chi}\xspace}
\def \mmm {\ensuremath{m(\mu^+\mu^-)}\xspace}
\def \tmm {\ensuremath{\tau(\mu^+\mu^-)}\xspace}
\def \mx {\ensuremath{m(\chi)}\xspace}
\def \tx {\ensuremath{\tau(\chi)}\xspace}
\def\jpsi     {{\ensuremath{{J\mskip -3mu/\mskip -2mu\psi\mskip 2mu}}}\xspace}
\def \mmmsq {\ensuremath{m^2(\mu^+\mu^-)}\xspace}

\makeatletter
\g@addto@macro\@floatboxreset\centering
\makeatother
\graphicspath{{./figs/}} 
\input{lhcb-symbols-def}

\begin{document}

\renewcommand{\thefootnote}{\fnsymbol{footnote}}
\setcounter{footnote}{1}


\begin{tabular*}{\linewidth}{lc@{\extracolsep{\fill}}r}
\ifthenelse{\boolean{pdflatex}}
 & & 11th of December 2015 \\
 & & \\
\end{tabular*}


{\bf\boldmath\huge
\begin{center}
Searches for Light Exotics at LHCb
\end{center}
}


\begin{center}
Federico Leo Redi\footnote{Imperial College London.},~on behalf of the LHCb Collaboration.
\end{center}


\begin{abstract}
  \noindent We report on the latest direct searches for light exotics at LHCb, conducted during  Run I of LHC.
  This proceedings are divided into two sections, the first part will cover the search for the lepton number violating decay $\decay{B}{\pi^{+}\mu^{-}\mu^{-}}$ while the second part will cover the search for a low mass dark boson in the decay $B^0\!\to K^{*0}\chi$, with $\chi \! \to \mu^+\mu^-$ and $K^{*0} \!\to K^+\pi^-$.
  The data used in these searches correspond to integrated luminosities of 1.0 and $2.0\fb^{-1}$ collected in $pp$ collisions at centre of mass energies of $\sqrt{s}=7$ and
  8 TeV in $pp$ collisions with the LHCb detector.
\end{abstract}




{\footnotesize
}







\section{Introduction}
The LHCb detector is one of the four main detectors that operate at the Large Hadron Collider (LHC) at CERN.
The LHCb detector consists of a single-arm forward spectrometer operating in the region of pseudorapidity, $1.9<\eta<4.9$.
The detector was originally designed to study the production and decay of hadrons containing $b$ and $c$ quarks and indirectly probing the strength of the Standard Model (SM).
The LHCb detector is now playing a fundamental role also in other areas of research, such as direct searches of rare SM decays and exotica.
Exotica searches at LHCb primarily consist  in Higgs physics and direct searches for beyond the SM particles.
Many theoretical models predict the existence of new particles:  their existence can be detected either directly through the production of on-shell particles or indirectly through virtual contributions in loop processes.

During Run I of the LHC, LHCb recorded data at a centre of mass energy $\sqrt{(s)}=7\tev$ (for 2010 and 2011) and $8\tev$ (for 2012) corresponding to an integrated luminosity of 1.0 and $2.0\fb^{-1}$ respectively.

The SM is an incomplete theory.
Not only the SM is in conflict with the observations of non-zero neutrino masses, the excess of matter over antimatter in the Universe, and the presence of non-baryonic dark matter but it also presents a number of fine-tuning problems (such as the hierarchy and strong CP problems).
Beyond the SM (BSM) physics has been searched for at the LHC without success so far.
Nevertheless LHCb is an ideal experiment to probe  unique regions BSM phase space thanks to the detector unique particle identification capabilities, precise secondary vertex reconstruction and accurate measurements of lifetime, momentum and invariant mass.
Throughout this document charge conjugation is implied unless explicitly stated otherwise and $c = 1$.

\section{Search for Majorana Neutrinos in $\decay{B}{\pi^{+}\mu^{-}\mu^{-}}$ Decays at LHCb}
The nature of neutrinos in the SM has not been defined yet: neutrinos could either be Dirac fermions or their own antiparticle. In the latter case they are called ``Majorana" particles~\cite{Majorana}.
Some of the most economical theories that can account simultaneously for neutrino masses and oscillations, baryogenesis, and dark matter, extend the SM by requiring the existance of a fourth neutrino generation.
Since a fourth neutrino generation can couple with SM particles there exist many ways of searching for such particles, one of them being the neutrino-less double $\beta$ decay.
The approach followed by LHCb is different and complementary, which performs a direct search in heavy flavour decays, similar to what has been done in the past~\cite{Aaij:2012zr,Aaij:2011ex,Lees:2013pxa,Seon:2011ni}.

The LHCb experiment has performed many studies for Majorana Neutrino produced in $B^{-}$ decays, probing a wide range of masses and lifetimes; these searches were performed  for the  lepton flavour violating decays  $B^-\to h^+\mu^-\mu^-$, where $h$ is a hadron.
These types of decays are prohibited by the SM but can happen thanks to production of on-shell Majorana neutrinos.
The LHCb collaboration published three papers using different final states and different data sets:
\begin{itemize}
  \item $h^+ = K^+$ or $\pi^+$, with $\sim$36 pb$^{-1}$ ($\sqrt{s}=$7 TeV) \cite{LHCb-PAPER-2011-009}.
  \item $h^+ = D^+$, $D^{\ast +}$, $D^+_s$ and $D^0 \pi^+$, with $\sim$410 pb$^{-1}$ ($\sqrt{s}=$7 TeV) \cite{LHCb-PAPER-2011-038}.
  \item $h^+ = \pi^+$, with 3.0 fb$^{-1}$ ($\sqrt{s}=$7 TeV + $\sqrt{s}=$8 TeV)~\cite{LHCb-PAPER-2013-064}.
\end{itemize}
This proceedings will concentrate on the latter paper, being the most recent of the three.

\begin{figure}
  \includegraphics[height=0.2\textheight]{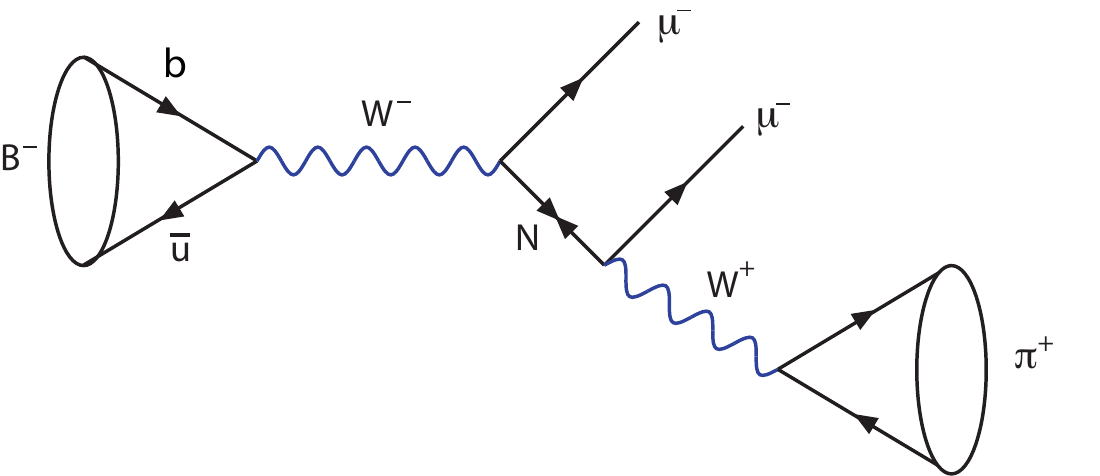}
  \caption{Feynman diagram for $\Bm\to\pi^+\mu^-\mu^-$ decay mediated by a Majorana neutrino ($N$). Reproduced from~\cite{LHCb-PAPER-2013-064}.}
  \label{fig:pimumu}
\end{figure}

A Feynman diagram for the lepton number and  flavour violating decay $B^-\to\pi^+\mu^-\mu^-$ is shown in Fig.~\ref{fig:pimumu}. This decay is prohibited by the SM but can happen thanks to production of on-shell Majorana neutrinos, it has been chosen as it is one of the most sensitive way to look for Majorana neutrinos in $B$ decays~\cite{LHCb-PAPER-2013-064}.
This decay, which has been theoretically modelled in Ref.~\cite{Atre:2009rg}, is sensitive to contributions from both on- and off-shell Majorana neutrino. More specifically if the mass of the Majorana neutrino, $m_N$, is smaller than $m_B - m_\mu$ then it can  be produced on-shell with a finite lifetime in the detector.
If, on the other hand, $m_N$ is larger then it can still contribute to the decay as a virtual particle.

The selection is  designed to maximise the efficiency squared divided by the background yield.
This allows for decay products to be detached from the  $B^-$ decay vertex, therefore  $\tau_N$ can span from few picoseconds  up to $\sim 1000 \ps$.
Because for lifetimes $\sim1 \ps$, the $\pi^+\mu^-$ vertex can be significantly detached from the $B^-$ decay vertex two different strategies are used: one for short $\tau_N$ ($\cal{S}$) and another for $\tau_N$  up to 1000 ps ($\cal{L}$).

In order to reduce the systematic uncertainty and to convert the yield into a branching fraction, the normalisation channel $B^- \rightarrow \jpsi K^-$ (with $\jpsi \to \mu^+\mu^-$) was chosen.
For the $\cal{S}$ category and the normalisation channel the $\mu^-\mu^-\pi^+$ candidate combinations must, when reconstructed, form a common vertex.
For the $\cal{L}$ category the $\pi^+\mu^-$ pair can be significantly displaced from the $\Bm$ vertex.
A \Bm candidate decay vertex is searched for by tracing back a neutrino $N$ candidate to another $\mu^{-}$ in the event, which must form a vertex.

\begin{figure}
  \includegraphics[width=1.\textwidth]{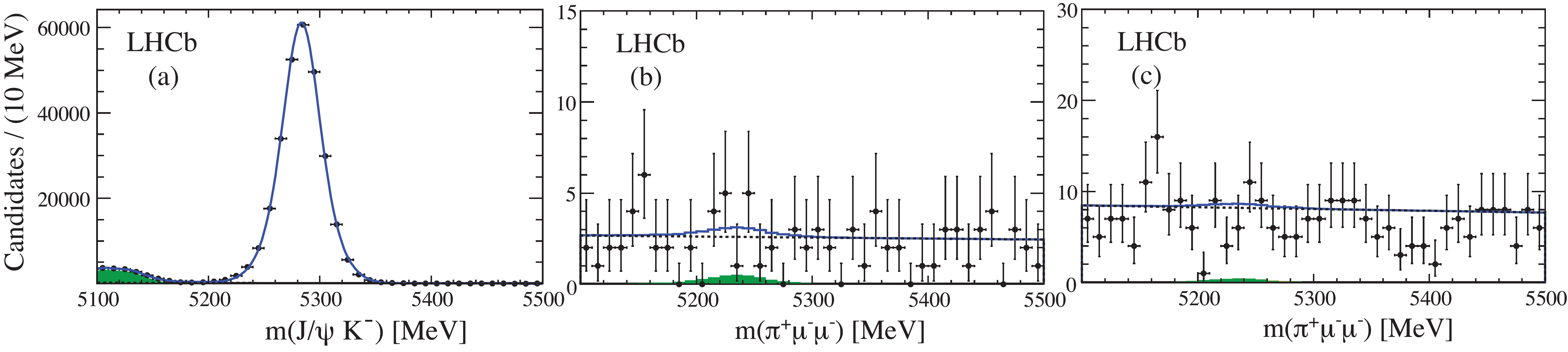}
  \caption{Invariant mass distributions with fits overlaid of candidate mass spectra for  (a) $\jpsi K^-$, (b) $\pip\mun\mun$ ($\cal{S}$) , and (c) $\pip\mun\mun$ ($\cal{L}$).   Where $\cal{S}$ $\cal{L}$ indicates the two different data samples  one for short $\tau_N$ and another for $\tau_N$  up to 1000 ps.
    Peaking backgrounds are (green) shaded. 
    The dotted lines show the combinatorial backgrounds only. The solid line shows the sum of both backgrounds~\cite{LHCb-PAPER-2013-064}.}
  \label{fig:mass}
\end{figure}

Figure~\ref{fig:mass} shows the mass spectra for the selected candidates.
No signal is observed in both the ${\cal{S}}$ and ${\cal{L}}$ samples.

\begin{figure}
\includegraphics[height=0.2\textheight]{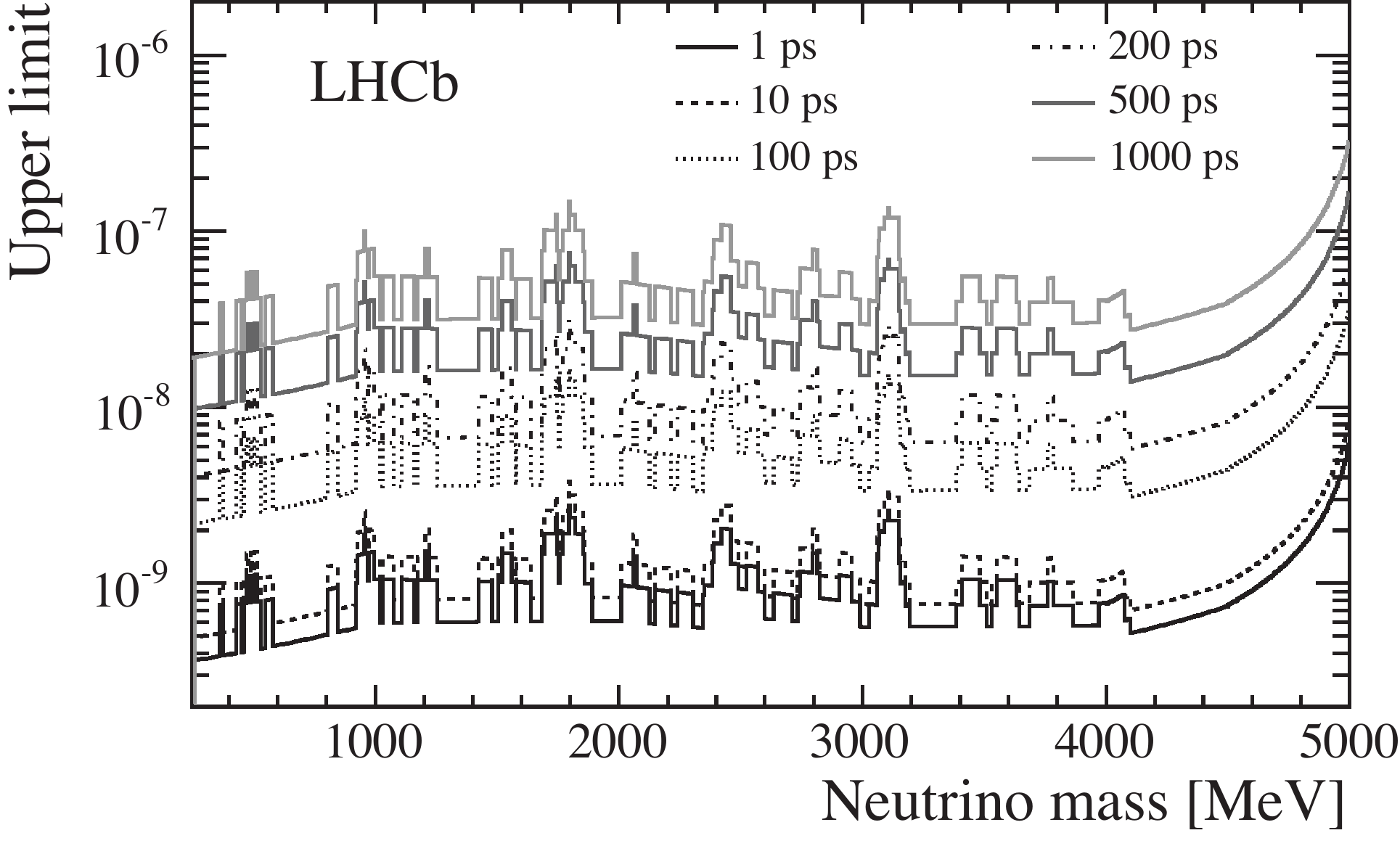}
\caption{Upper limits on ${\cal{B}}(\Bm \to \pip \mun \mun)$ at 95\% C.L.  as a function of $m_N$,  in 5~MeV intervals, for specific values of $\tau_N$~\cite{LHCb-PAPER-2013-064}.}
\label{fig:UL-mass-lifetime}
\end{figure}

In order to set upper limits the Confidence Level method is used~\cite{CLS}.
The signal region is defined as the mass interval within $\pm 2\sigma$ of the \Bm mass where $\sigma$ is the mass resolution.
Because no evidence for a signal is found, upper limits are set by scanning across the $m_N$ window.
The efficiency is highest for $\tau_N$ of a few ps, then it decreases rapidly until $\tau_N \sim200 \ps$ when it levels off until $\tau_N\sim 1000 \ps$.
After this value, the efficiency decreases to $\sim 0$ because most of the decays happen outside of the vertex detector.
The multi-dimensional plot of the upper limit on ${\cal{B}}(\Bm \to \pip \mun \mun)$ is shown in Fig.~\ref{fig:UL-mass-lifetime}.

\begin{figure}
\includegraphics[height=0.2\textheight]{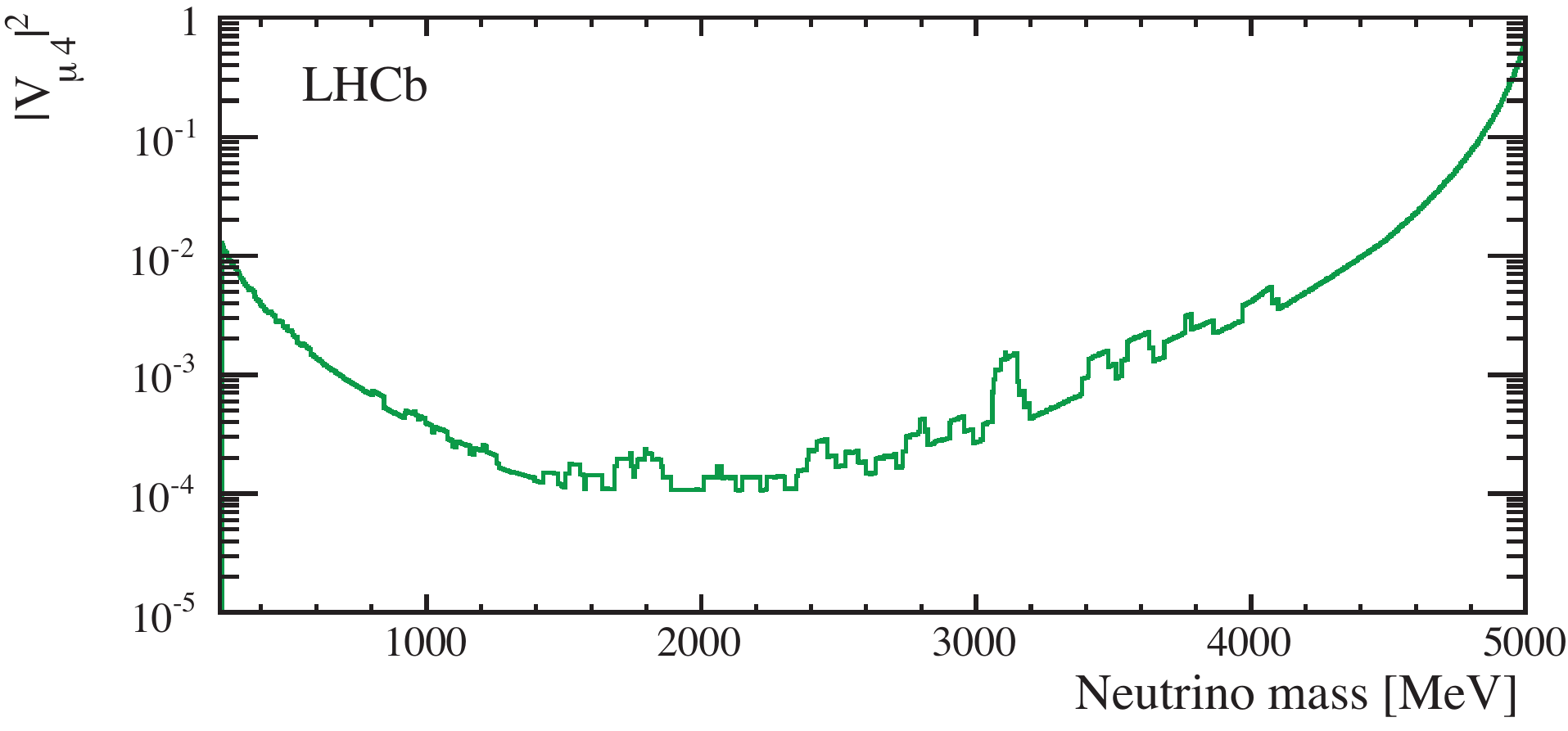}
\caption{Upper limits at 95\% C.L. on the fourth generation neutrino coupling to the muon $|V_{\mu4}|^2$ are shown as a function of the mass of the Majorana neutrino  $m_N$ for the events in the displaced region~\cite{LHCb-PAPER-2013-064}.}
\label{fig:Vmu4modified}
\end{figure}

A model dependent upper limit on the coupling of a single fourth-generation Majorana neutrino to muons, $|V_{\mu 4}|$, for each value of $m_N$,  is calculated using an expansion of the formula used by Ref.~\cite{Atre:2009rg}.
The resulting 95\% C.L. limit on $|V_{\mu4}|^2$ is extracted as a function of  $m_N$ and is shown in Fig.~\ref{fig:Vmu4modified}.

\section{A Search for the Decay of a Hidden Sector Particle $\decay{\chi}{\mu{+}\mu^{-}}$ in $\decay{B^{0}}{K^{*0}\chi}$ at LHCb}
In particle physics, the term hidden-sector refers to the set of predicted particles that do not interact via the gauge boson forces of the SM.
Interest in hidden-sector SM extensions has increased~\cite{Essig:2013lka}  due to the lack of any new TeV scale particles and missing evidence for a dark matter candidate that could solve the open questions in high energy physics~\cite{Weidenspointner:2006nua,Chang:2008aa,Adriani:2008zr,Adriani:2011xv,Adriani:2013uda,FermiLAT:2011ab,Aguilar:2014mma}.
As for the Majorana neutrino, coupling between the SM and hidden-sector particles may arise via mixing between the hidden-sector field and any SM field with an associated particle that is not charged under the electromagnetic or strong interaction.
This mixing could  provide a portal through which a hidden-sector particle, $\chi$, may be produced when kinematically allowed.
This proceedings will concentrate on the search performed by LHCb for a hidden-sector boson produced in the decay $B^0\!\to K^{*0}\chi$, with $\chi \! \to \mu^+\mu^-$ and $K^{*0} \!\to K^+\pi^-$ (throughout this proceedings, $K^{*0} \equiv K^{*}(892)^0$)~\cite{LHCb-PAPER-2015-036}.

\begin{figure}
  \includegraphics[height=0.2\textheight]{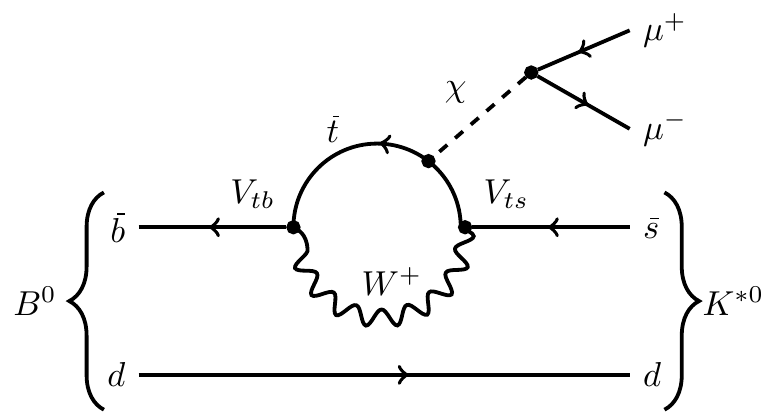}
  \caption{
    Feynman diagram for the decay \sigdecay,  with $\chi \! \to \mu^+\mu^-$~\cite{LHCb-PAPER-2015-036}.
    \label{fig:feyDB}}
\end{figure}

As shown in Fig.~\ref{fig:feyDB},   the $b\!\to s$ transition is mediated by a top quark loop at leading order.
For this reason, $\chi$ boson with a sizeable top quark coupling, \eg obtained via mixing with the Higgs sector, could be produced at a substantial rate in such decays.
The dataset used for this analysis is the same used for the Majorana neutrino search reported in the previous chapter.

The  search is conducted, as outlined in Ref.~\cite{Williams:2015xfa}, by scanning the \mmm distribution for an excess of $\chi$ signal candidates over the expected background.
The $\chi\!\to\mu^+\mu^-$ decay vertex is permitted, but not required, to be displaced from the \sigdecay decay vertex.  Two
regions of reconstructed dimuon lifetime, \tmm, are defined for each \mx considered in the search: a prompt region and a displaced region.
Narrow resonances are vetoed by excluding the regions near the $\omega$, $\phi$, $J/\psi$, $\psi(2S)$ and $\psi(3770)$ resonances.  These regions are removed in both the prompt and displaced samples to avoid contamination from unassociated dimuon and $K^{*0}$ resonances.

\begin{figure}
  \includegraphics[height=0.2\textheight]{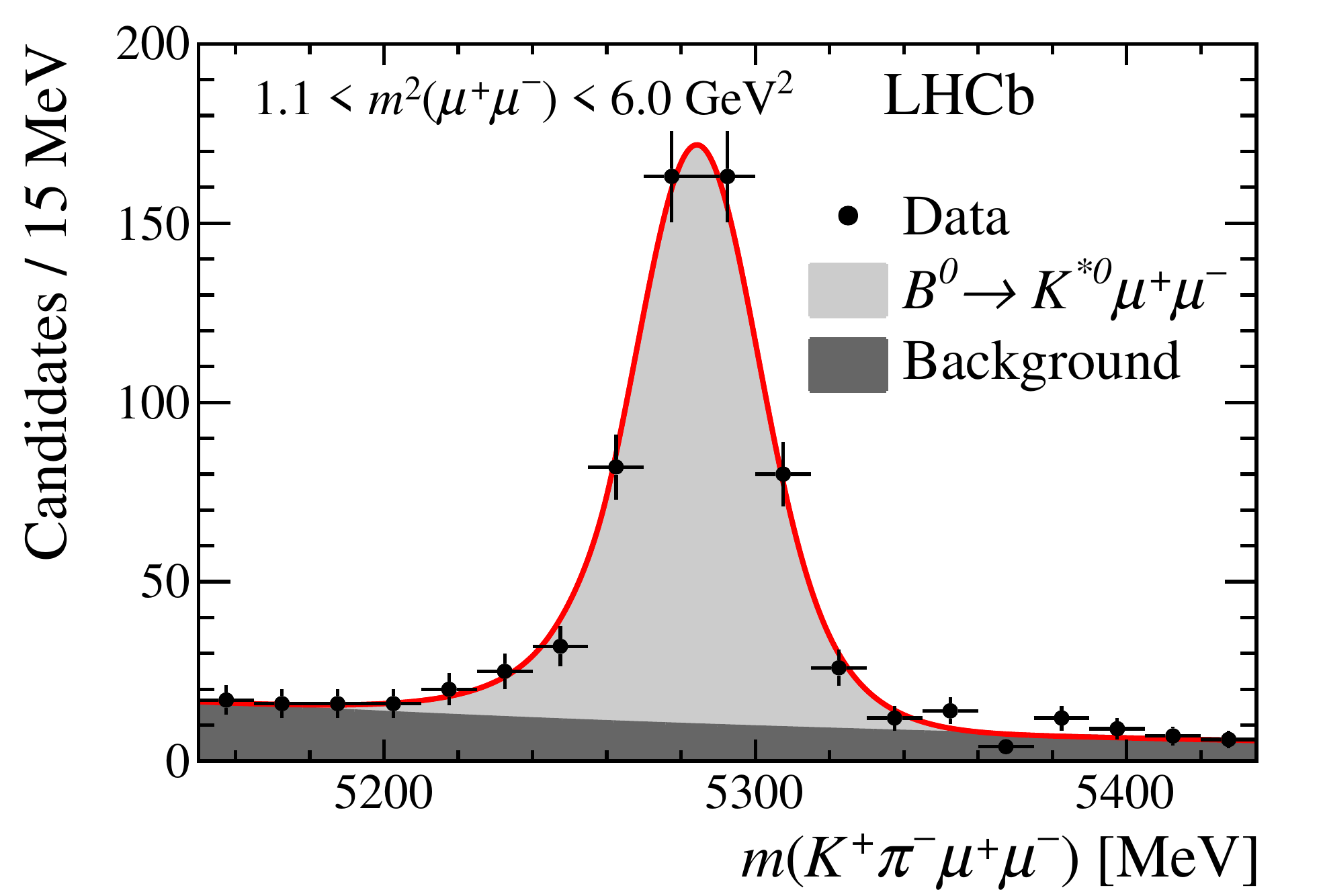}
  \caption{
    Invariant mass spectrum with fit overlaid~\cite{LHCb-PAPER-2015-036}. 
    \label{fig2}}
\end{figure}

The branching fraction product $\mathcal{B}(B^0\!\to K^{*0}\chi(\mu^+\mu^-))\equiv\mathcal{B}(\sigdecay)\times\mathcal{B}(\chi\!\to\mu^+\mu^-)$ is measured relative to $\mathcal{B}(\smdecay)$, where the normalisation sample is taken from the prompt region.
Figure~\ref{fig2} shows the $K^+\pi^-\mu^+\mu^-$ control channel mass distribution for all prompt candidates that satisfy the full selection.
An extended unbinned likelihood fit is performed on the control channel to the mass spectrum.

\begin{figure}
  \includegraphics[height=0.15\textheight]{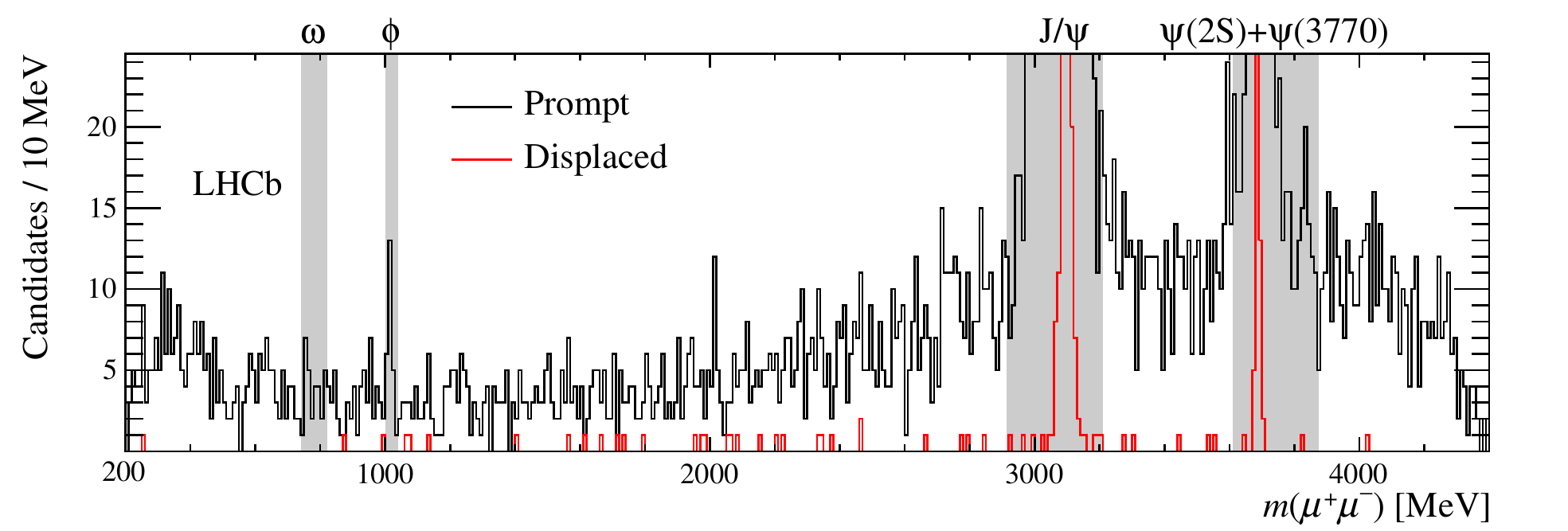}
  \caption{
    Distribution of \mmm in the (black) prompt and (red) displaced regions.  The shaded bands denote regions where no search is performed due to (possible) resonance contributions.  The \jpsi, $\psi(2S)$ and $\psi(3770)$ peaks are suppressed to better display the search region~\cite{LHCb-PAPER-2015-036}.
    \label{fig3}}
\end{figure}

\begin{figure}
  \includegraphics[height=0.15\textheight]{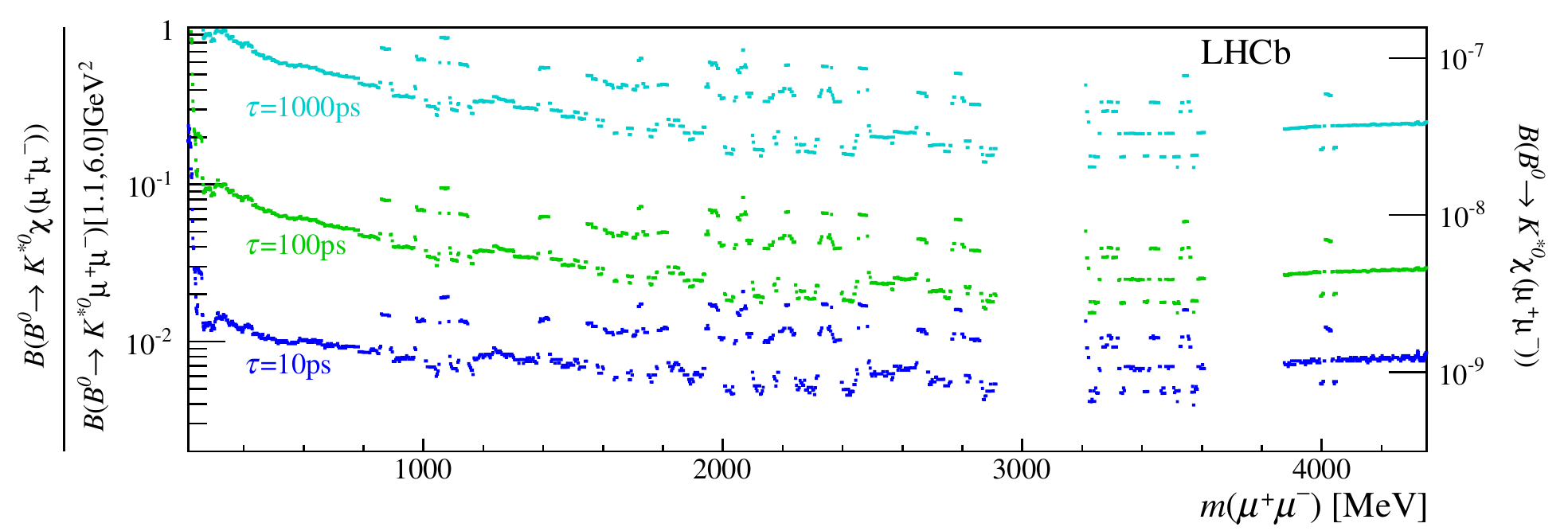}
  \caption{
    Upper limits at 95\% CL for (left axis)
    $\mathcal{B}(\sigdecay(\mu^+\mu^-))/\mathcal{B}(\smdecay)$, with
    \smdecay in $1.1 < \mmmsq < 6.0\gev^2$, and (right axis)
    $\mathcal{B}(\sigdecay(\mu^+\mu^-))$~\cite{LHCb-PAPER-2015-036}.
    \label{fig4}}
\end{figure}

The \mmm distributions in both the prompt and displaced regions for candidates with an invariant mass that lies in a window of $50\mev$ around the known $B^0$ mass are shown in Fig.~\ref{fig3}.
The $p$-value of the no-signal hypothesis is  80\%, showing  no evidence for a hidden-sector boson.
Because no signal events are found, Fig.~\ref{fig4} shows the upper limits on $\mathcal{B}(\sigdecay(\mu^+\mu^-))$, relative to $\mathcal{B}(\smdecay)$, set at the 95\% C.L. for different values of \tx.
As the Figure shows, the limits become less stringent for $\tx \gtrsim 10\ps$, as the probability of the dark boson decaying within the vertex locator decreases.
The branching fraction $\mathcal{B}(\smdecay)=(1.6\pm0.3)\times10^{-7}$~\cite{LHCb-PAPER-2013-019} is used to obtain upper limits on $\mathcal{B}(\sigdecay(\mu^+\mu^-))$, which are also shown in  Fig.~\ref{fig4}.

\section{Conclusions}
\begin{figure}
 \includegraphics[height=0.2\textheight]{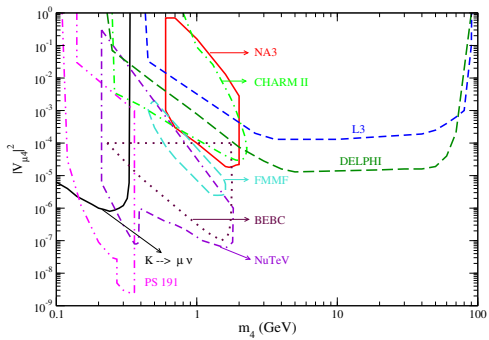}
 \caption{Limits on $|V_{\mu 4}|$ as a function of $m_4$ as set by previous non LHCb results. The area indicated by dotted lines are at 95\% confidence level. Reproduced from~\cite{Atre:2009rg}.}
  \label{fig:boundsUmu}
\end{figure}

In these proceedings we have provided two examples of direct searches for light exotics in the LHCb detector.
This shows how LHCb can make contributions in the intensity frontier searches for  BSM physics,  where light particles are rarely coupling to the SM filed.
As an example, Fig.~\ref{fig:boundsUmu} shows the existing experimental limits for the mixing parameter  $|V_{\mu 4}|$ as a function of the Majorana neutrino candidate mass.
It is striking that DELPHI was the last experiment to set a limit in the region of phase space above the charm quark mass.
This means that   LHCb  is one of the few experiments, up to date, able to further constrain the phase space for this parameter.

\addcontentsline{toc}{section}{References}
\setboolean{inbibliography}{true}
\bibliographystyle{LHCb}
\bibliography{paper,LHCB-PAPER}
\end{document}

%% file: lhcb-symbols-def.tex
\usepackage{ifthen}
\newboolean{uprightparticles}
\setboolean{uprightparticles}{false} 
\usepackage{xspace}
\usepackage{upgreek}

\def\MagUp {\mbox{\em Mag\kern -0.05em Up}\xspace}

\ifthenelse{\boolean{uprightparticles}}%
{

 \def\Pmu         {\ensuremath{\upmu}\xspace}

 \def\Ppi         {\ensuremath{\uppi}\xspace}

 \def\Ppsi        {\ensuremath{\uppsi}\xspace}

 \def\PDelta      {\ensuremath{\Delta}\xspace}
 \def\PXi      {\ensuremath{\Xi}\xspace}
 \def\PLambda      {\ensuremath{\Lambda}\xspace}
 \def\PSigma      {\ensuremath{\Sigma}\xspace}
 \def\POmega      {\ensuremath{\Omega}\xspace}
 \def\PUpsilon      {\ensuremath{\Upsilon}\xspace}

 \def\PB      {\ensuremath{\mathrm{B}}\xspace}
 
 \def\PD      {\ensuremath{\mathrm{D}}\xspace}

 \def\PJ      {\ensuremath{\mathrm{J}}\xspace}
 \def\PK      {\ensuremath{\mathrm{K}}\xspace}

 \def\Pi      {\ensuremath{\mathrm{i}}\xspace}

}
{

 \def\Pmu         {\ensuremath{\mu}\xspace}

 \def\Ppi         {\ensuremath{\pi}\xspace}

 \def\Ppsi        {\ensuremath{\psi}\xspace}
 
 \mathchardef\PDelta="7101
 \mathchardef\PXi="7104
 \mathchardef\PLambda="7103
 \mathchardef\PSigma="7106
 \mathchardef\POmega="710A
 \mathchardef\PUpsilon="7107
 
 \def\PB      {\ensuremath{B}\xspace}
 
 \def\PD      {\ensuremath{D}\xspace}

 \def\PJ      {\ensuremath{J}\xspace}
 \def\PK      {\ensuremath{K}\xspace}

 \def\Pi      {\ensuremath{i}\xspace}

}

\makeatletter
\ifcase \@ptsize \relax
  \newcommand{\miniscule}{\@setfontsize\miniscule{4}{5}}
\or
  \newcommand{\miniscule}{\@setfontsize\miniscule{5}{6}}
\or
  \newcommand{\miniscule}{\@setfontsize\miniscule{5}{6}}
\fi
\makeatother

\DeclareRobustCommand{\optbar}[1]{\shortstack{{\miniscule (\rule[.5ex]{1.25em}{.18mm})}
  \\ [-.7ex] $#1$}}

\def\mun        {{\ensuremath{\Pmu^-}}\xspace} 

\def\pion   {{\ensuremath{\Ppi}}\xspace}

\def\pip    {{\ensuremath{\pion^+}}\xspace}

  \def\Kbar    {{\kern 0.2em\overline{\kern -0.2em \PK}{}}\xspace}

\def\KorKbar    {\kern 0.18em\optbar{\kern -0.18em K}{}\xspace}

  \def\Dbar    {{\kern 0.2em\overline{\kern -0.2em \PD}{}}\xspace}

\def\DorDbar    {\kern 0.18em\optbar{\kern -0.18em D}{}\xspace}

\def\B       {{\ensuremath{\PB}}\xspace}
\def\Bbar    {{\ensuremath{\kern 0.18em\overline{\kern -0.18em \PB}{}}}\xspace}

\def\BorBbar    {\kern 0.18em\optbar{\kern -0.18em B}{}\xspace}

\def\Bub     {{\ensuremath{\B^-}}\xspace}

\def\Bm      {{\ensuremath{\Bub}}\xspace}

\def\jpsi     {{\ensuremath{{\PJ\mskip -3mu/\mskip -2mu\Ppsi\mskip 2mu}}}\xspace}

  \def\Y#1S{\ensuremath{\PUpsilon{(#1S)}}\xspace}

\def\Lbar        {{\ensuremath{\kern 0.1em\overline{\kern -0.1em\PLambda}}}\xspace}
\def\LorLbar    {\kern 0.18em\optbar{\kern -0.18em \PLambda}{}\xspace}

\newcommand{\decay}[2]{\ensuremath{#1\!\to #2}\xspace}         

\def\to                 {\ensuremath{\rightarrow}\xspace}

\def\AT#1     {\ensuremath{A_{\mathrm{T}}^{#1}}\xspace}           

\def\C#1      {\ensuremath{\mathcal{C}_{#1}}\xspace}                       
\def\Cp#1     {\ensuremath{\mathcal{C}_{#1}^{'}}\xspace}                    
\def\Ceff#1   {\ensuremath{\mathcal{C}_{#1}^{\mathrm{(eff)}}}\xspace}        
\def\Cpeff#1  {\ensuremath{\mathcal{C}_{#1}^{'\mathrm{(eff)}}}\xspace}       
\def\Ope#1    {\ensuremath{\mathcal{O}_{#1}}\xspace}                       
\def\Opep#1   {\ensuremath{\mathcal{O}_{#1}^{'}}\xspace}                    

\newcommand{\tev}{\ensuremath{\mathrm{\,Te\kern -0.1em V}}\xspace}
\newcommand{\gev}{\ensuremath{\mathrm{\,Ge\kern -0.1em V}}\xspace}
\newcommand{\mev}{\ensuremath{\mathrm{\,Me\kern -0.1em V}}\xspace}
\newcommand{\kev}{\ensuremath{\mathrm{\,ke\kern -0.1em V}}\xspace}
\newcommand{\ev}{\ensuremath{\mathrm{\,e\kern -0.1em V}}\xspace}
\newcommand{\gevc}{\ensuremath{{\mathrm{\,Ge\kern -0.1em V\!/}c}}\xspace}
\newcommand{\mevc}{\ensuremath{{\mathrm{\,Me\kern -0.1em V\!/}c}}\xspace}
\newcommand{\gevcc}{\ensuremath{{\mathrm{\,Ge\kern -0.1em V\!/}c^2}}\xspace}
\newcommand{\gevgevcccc}{\ensuremath{{\mathrm{\,Ge\kern -0.1em V^2\!/}c^4}}\xspace}
\newcommand{\mevcc}{\ensuremath{{\mathrm{\,Me\kern -0.1em V\!/}c^2}}\xspace}

\def\fb   {\ensuremath{\mbox{\,fb}}\xspace}

\def\ps   {\ensuremath{{\mathrm{ \,ps}}}\xspace}

\def\gsim{{~\raise.15em\hbox{$>$}\kern-.85em
          \lower.35em\hbox{$\sim$}~}\xspace}
\def\lsim{{~\raise.15em\hbox{$<$}\kern-.85em
          \lower.35em\hbox{$\sim$}~}\xspace}

\def\tell1  {TELL1\xspace}
\def\ukl1   {UKL1\xspace}

\newcommand{\eg}{\mbox{\itshape e.g.}\xspace}